\documentclass[twocolumn,english,aps,prb,twocolum,superscriptaddress,bibnotes,amsmath,amssymb,floatfix]{revtex4-1}
\usepackage[colorlinks=true,citecolor=blue,linkcolor=magenta]{hyperref}

\usepackage[markup=nocolor, authormarkupposition=left]{changes} 
\usepackage{soul}
\usepackage[utf8]{inputenc}
\usepackage[english]{babel}
\usepackage{amsmath,amsfonts,amssymb}
\usepackage[T1]{fontenc}
\usepackage{url}

\usepackage{amsmath}
\usepackage{amsfonts}
\usepackage{amssymb}

\usepackage{epstopdf}
\usepackage{graphicx}
\graphicspath{{./Figures/}}

\begin{document}
\title{Monolithic piezoelectric control of soliton microcombs}

\author{Junqiu Liu}
\thanks{These authors contributed equally to this work.}
\affiliation{Institute of Physics, Swiss Federal Institute of Technology Lausanne (EPFL), CH-1015 Lausanne, Switzerland}

\author{Hao Tian}
\thanks{These authors contributed equally to this work.}
\affiliation{OxideMEMS Lab, Purdue University, 47907 West Lafayette, IN, USA}

\author{Erwan Lucas}
\thanks{These authors contributed equally to this work.}
\affiliation{Institute of Physics, Swiss Federal Institute of Technology Lausanne (EPFL), CH-1015 Lausanne, Switzerland}

\author{Arslan S. Raja}
\thanks{These authors contributed equally to this work.}
\affiliation{Institute of Physics, Swiss Federal Institute of Technology Lausanne (EPFL), CH-1015 Lausanne, Switzerland}

\author{Grigory Lihachev}
\affiliation{Institute of Physics, Swiss Federal Institute of Technology Lausanne (EPFL), CH-1015 Lausanne, Switzerland}

\author{Rui Ning Wang}
\affiliation{Institute of Physics, Swiss Federal Institute of Technology Lausanne (EPFL), CH-1015 Lausanne, Switzerland}

\author{Jijun He}
\affiliation{Institute of Physics, Swiss Federal Institute of Technology Lausanne (EPFL), CH-1015 Lausanne, Switzerland}

\author{Tianyi Liu}
\affiliation{Institute of Physics, Swiss Federal Institute of Technology Lausanne (EPFL), CH-1015 Lausanne, Switzerland}

\author{Miles H. Anderson}
\affiliation{Institute of Physics, Swiss Federal Institute of Technology Lausanne (EPFL), CH-1015 Lausanne, Switzerland}

\author{Wenle Weng}
\affiliation{Institute of Physics, Swiss Federal Institute of Technology Lausanne (EPFL), CH-1015 Lausanne, Switzerland}

\author{Sunil A. Bhave}
\email[]{bhave@purdue.edu}
\affiliation{OxideMEMS Lab, Purdue University, 47907 West Lafayette, IN, USA}

\author{Tobias J. Kippenberg}
\email[]{tobias.kippenberg@epfl.ch}
\affiliation{Institute of Physics, Swiss Federal Institute of Technology Lausanne (EPFL), CH-1015 Lausanne, Switzerland}

\maketitle

\noindent\textbf{
High-speed laser frequency actuation is critical in all applications employing lasers and frequency combs, and is prerequisite for phase locking, frequency stabilization and stability transfer among multiple optical carriers.
Soliton microcombs\cite{Kippenberg:18, Gaeta:19} have emerged as chip-scale, broadband and low-power-consumption frequency comb sources.
Yet, integrated microcombs relying on thermal heaters\cite{Joshi:16, Xue:16} for on-chip actuation all exhibit only kilohertz actuation bandwidth.
Consequently, high-speed actuation and locking of microcombs have been attained only with off-chip bulk modulators.
Here, we present high-speed microcomb actuation using integrated components.
By monolithically integrating piezoelectric AlN actuators on ultralow-loss Si$_3$N$_4$ photonic circuits, we demonstrate voltage-controlled soliton tuning, modulation and stabilization.
The integrated AlN actuators feature bi-directional tuning with high linearity and low hysteresis, operate with 300 nW power and exhibit flat actuation response up to megahertz frequency, significantly exceeding bulk piezo tuning bandwidth. 
We use this novel capability to demonstrate a microcomb engine for parallel FMCW LiDAR\cite{Riemensberger:19, Kuse:19}, via synchronously tuning the laser and microresonator. 
By applying a triangular sweep at the modulation rate matching the frequency spacing of HBAR modes\cite{Tian:19}, we exploit the resonant build-up of bulk acoustic energy to significantly lower the required driving to a CMOS voltage of only 7 Volts. 
Our approach endows soliton microcombs with integrated, ultralow-power-consumption, and fast actuation, significantly expanding the repertoire of technological applications.
}

\begin{figure*}[t!]
\centering
\includegraphics{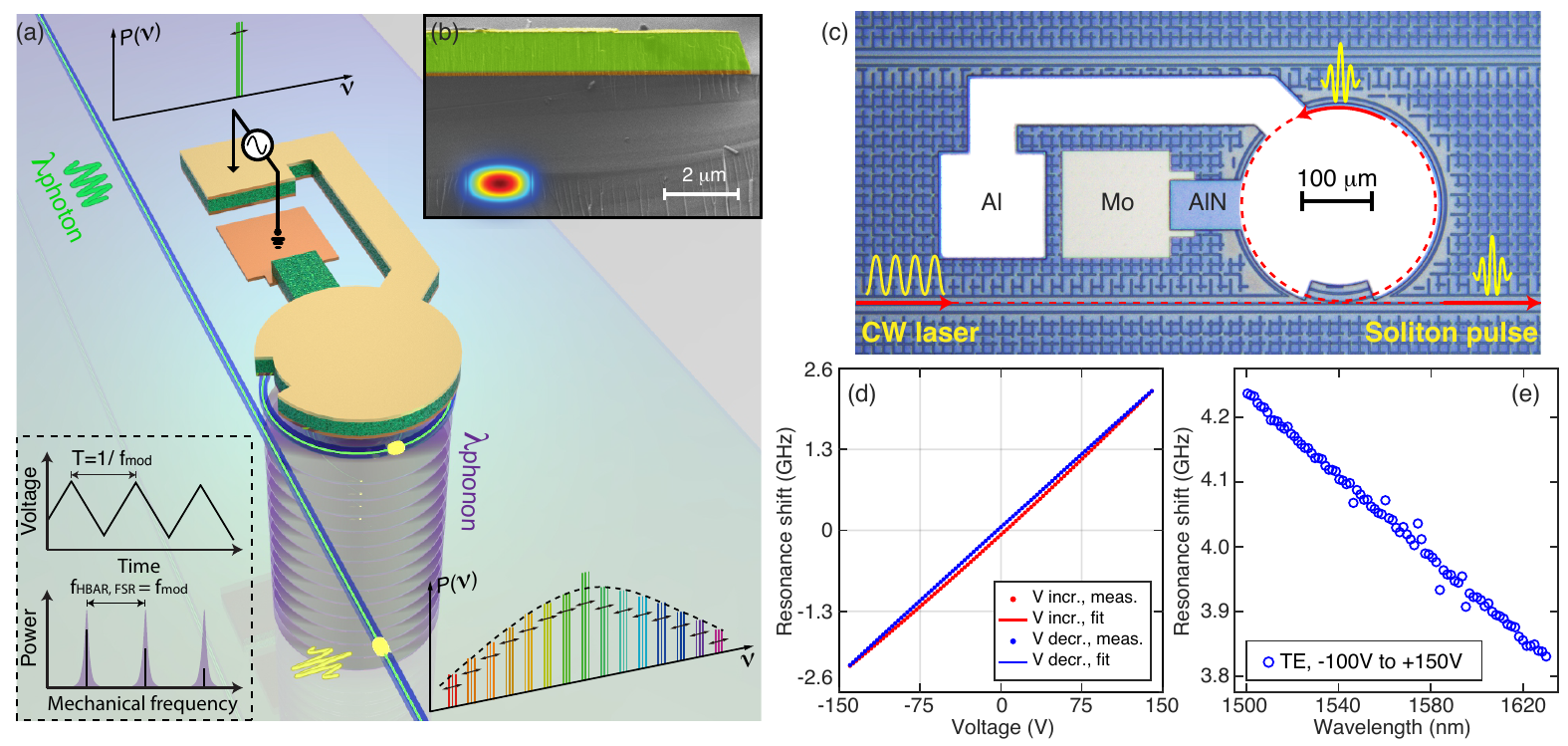}
\caption{
\textbf{Principle of monolithic piezoelectric control of chip-based soliton microcombs}.
(a)~Schematic of using piezoelectric AlN to actuate the soliton microcomb, via either stress-optic effect or HBAR modes.
Left-bottom inset: Fast and efficient microresonator modulation is achieved when applying a triangular signal whose Fourier frequency components match the HBAR modes.
(b)~False-colored SEM image of the sample cross-section, showing Al (yellow), AlN (green), Mo (red), and Si$_3$N$_4$ (blue).
(c)~Microscope image showing the Si$_3$N$_4$ microresonator with a disk-shape AlN actuator.
(d)~Resonance shift versus applied voltage. 
The linear tuning coefficient at DC is $\delta\nu/\delta V=15.7$~MHz/V. 
A weak hysteresis is observed.
(e)~Resonance shift for all TE$_{00}$ resonances in the wavelength range from 1500~nm to 1630~nm, when the applied voltage changes from $-100$~V to $+150$~V. 
The wavelength-dependent tuning results from $\sim5$~MHz change in microresonator FSR.
}
\label{Fig:1}
\end{figure*}

In recent years there has been major progress in soliton microcombs~\cite{Kippenberg:18} which constitute chip-scale, broadband, high-repetition-rate, coherent frequency combs.
These soliton microcombs are generated in nonlinear optical microresonators driven by continuous-wave (CW) lasers, in which dissipative Kerr solitons~\cite{HerrNP:14} are formed.
This formation process of ``dissipative structures'' is also found in other systems driven out of equilibrium.
Importantly, recent advances in CMOS-compatible low-loss photonic integrated circuits have enabled chip-based microcombs~\cite{Gaeta:19} that could allow co-integration of multiple functionalities on a chip, as well as compact, low-power-consumption, electronically controlled frequency comb devices~\cite{Stern:18, Raja:19, Shen:19} for field-deployable applications.

Silicon nitride (Si$_3$N$_4$)~\cite{Moss:13} has emerged as a leading platform for integrated soliton microcombs \cite{Gaeta:19}, 
based on which several system-level applications have been demonstrated, including coherent communication~\cite{Marin-Palomo:17}, distance measurement~\cite{Trocha:18, Suh:18}, astronomical spectrometer calibration~\cite{Suh:19, Obrzud:19}, as well as optical frequency synthesizers~\cite{Spencer:18} and atomic clocks~\cite{Newman:19}.
For many applications of frequency combs, such as optical frequency synthesis~\cite{Jones:00}, frequency division~\cite{Fortier:11}, and dual-comb spectroscopy~\cite{Coddington:16}, the capability to achieve megahertz-bandwidth actuation of comb teeth, as well as the repetition rate, is critical.
In contrast to femtosecond laser frequency combs which have developed a wide range of fast actuators within the laser cavity, measurement-based feedback stabilization of photonic chip-based microcombs still relies on off-chip, bulk acousto-optic modulators (AOM) or electro-optic modulators (EOM) that actuate on the pump laser. 
Therefore, high-speed actuators integrated on chip are highly desirable for microcombs.
While integrated modulators on Si$_3$N$_4$ waveguides have been demonstrated, e.g. based on the electro-absorption of graphene~\cite{Gruhler:13, Phare:15, Wang:15} and ferroelectric lead-zirconate-titanate (PZT)~\cite{Hosseini:15, Alexander:18, Jin:18}, these methods are not well suited for integrated microcomb applications. 
Graphene-based actuators are not compatible with wafer-scale manufacture using foundry processes, and ferroelectric PZT actuators exhibit tuning hysteresis which poses challenges for soliton initiation and switching.
Importantly, these methods have not yet demonstrated compatibility with soliton generation, which requires maintaining ultralow optical losses in Si$_3$N$_4$ waveguides. 
Therefore, current integrated actuation techniques for soliton microcombs rely primarily on metallic heaters~\cite{Joshi:16, Xue:16}, which have only kilohertz actuation bandwidth, limited by the thermal relaxation time.
In addition, heaters exhibit uni-directional tuning, and typically consume electrical power exceeding 30~mW \cite{Stern:18}, higher than the threshold optical power for soliton formation in state-of-the-art integrated devices~\cite{Stern:18, Raja:19, Shen:19}, and are not compatible with cryogenic operation~\cite{Moille:19, Stanfield:19}.
Alternatively, direct soliton generation in materials exhibiting the Pockels effect \cite{Li:19}, such as AlN~\cite{Xiong:12, LiuX:18}, LiNbO$_3$~\cite{Gong:19, He:19} and AlGaAs \cite{Pu:16, Chang:19}, could allow simultaneous high-speed actuation and soliton control, however these platforms yet are not as mature as Si$_3$N$_4$.

Here, we demonstrate integrated piezoelectric actuators which overcome these limitations. 
They are based on aluminium nitride (AlN)~\cite{Muralt:04}, a commercial micro-electro-mechanical-systems (MEMS) technology to build microwave filters in modern cellular phone technology~\cite{Dubois:99} and aeroacoustic microphones~\cite{Williams:12}.
We show that the piezoelectric control employing the stress-optic effect~\cite{Huang:03, vanderSlot:19} allows deterministic soliton initiation, switching, tuning, long-term stabilization, and phase locking with $\sim0.6$~MHz bandwidth.
Such integrated piezoelectric actuators that are linear, fast, non-absorptive, bi-directional, and consume ultralow electric power endow integrated $\mathrm{Si_3N_4}$ soliton microcombs with novel capability in power-critical applications e.g. in space, data centers and portable atomic clocks, in extreme environment such as cryogenic temperatures, or in emerging applications for coherent LiDAR~\cite{Kuse:19, Riemensberger:19}, frequency synthesizers~\cite{Spencer:18, Liu:19} and RF photonics~\cite{Torres-Company:14, Wu:18}.  

\begin{figure*}[t!]
\centering
\includegraphics{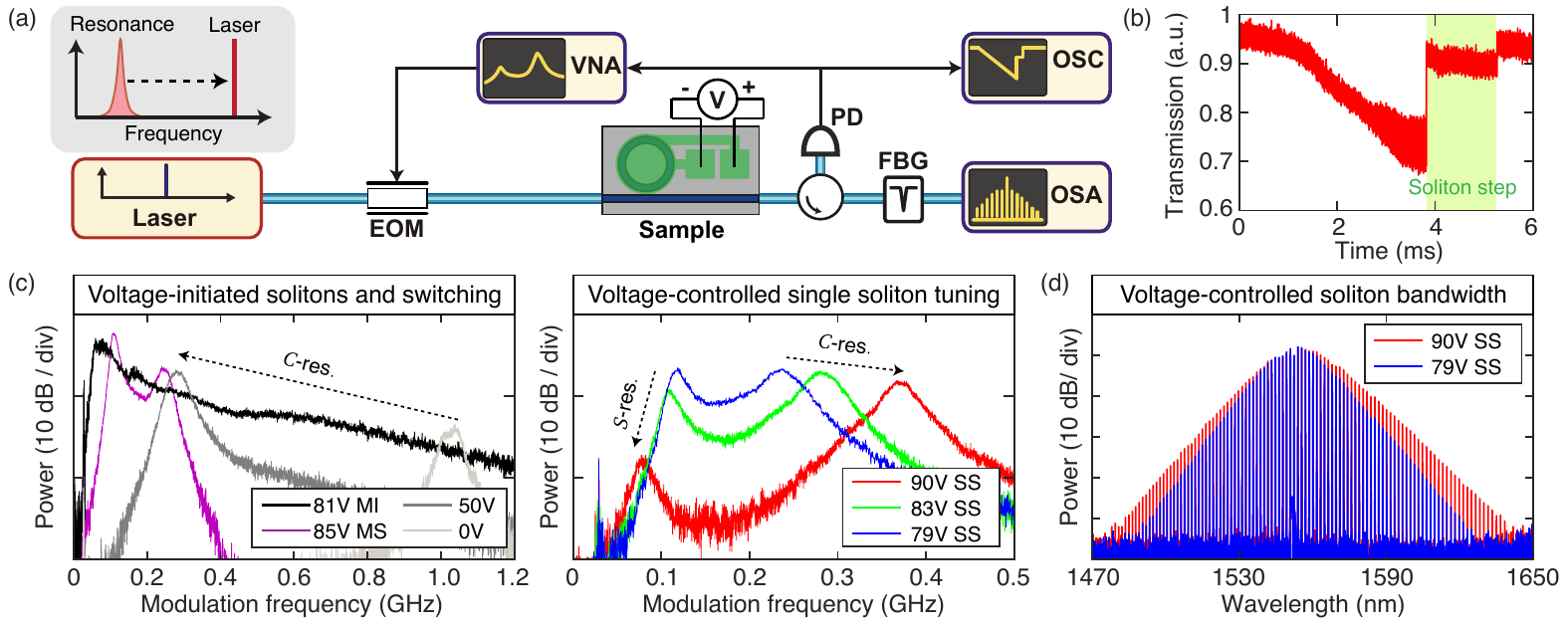}
\caption{
\textbf{Piezoelectric voltage-controlled soliton initiation and tuning}.
(a)~Experimental setup. 
(b)~A typical soliton step featuring millisecond length.
(c)~Soliton detuning control via AlN actuation. 
Left: Initially the resonance is 1~GHz blue-detuned from the laser (0~V). 
The resonance is tuned to the laser, and generate modulation instability (MI, 81~V) and a multi-soliton state (MS, 85~V). 
Right: Once the multi-soliton is generated (85~V), the voltage is reduced to switch to enable switching to a single soliton state (SS, 79~V). 
The soliton detuning, as well as the bandwidth, is further increased via increasing the voltage (90~V).
$\mathcal{S}$-res.: Soliton resonance. $\mathcal{C}$-res.: Cavity resonance. 
(d)~Different soliton states with different applied voltage. 
}
\label{Fig:2}
\end{figure*}

\textbf{Device fabrication and characterization}: 
Figure \ref{Fig:1}(a) shows the schematic of using the piezoelectric AlN to actuate the soliton microcomb, via either stress-optic effect or high-overtone bulk acoustic resonator (HBAR) modes.
The inset shows the scheme of efficient and fast microcomb actuation by modulating the microresonator with Fourier frequency components matching the HBAR modes\cite{Tian:19}, which will be discussed later.
The AlN actuators are monolithically integrated on ultralow-loss Si$_3$N$_4$ waveguides fabricated using the photonic Damascene reflow process\cite{Pfeiffer:18}.
In order to preserve the ultralow waveguide loss, a 2.4-$\mathrm{\mu}$m-thick SiO$_2$ top cladding is deposited on the Si$_3$N$_4$ waveguides, before the piezoelectric actuators are deposited and patterned.
The piezoelectric actuators\cite{Tian:19} are made from polycrystalline AlN as the main piezoelectric material, molybdenum (Mo) as the bottom electrode (ground) and the substrate to grow polycrystalline AlN, and aluminium (Al) as the top electrode.
The polycrystalline AlN has a piezoelectric coefficient\cite{Tsubouchi:85} $C_{33}\sim1.55$ C/m$^2$, and Mo is chosen in order to minimize the acoustic impedance with AlN\cite{Zhang:05} compared to gold or titanium.
Figure~\ref{Fig:1}(b) shows the scanning electron microscope (SEM) image of the sample cross-section, including the optical mode in the Si$_3$N$_4$ waveguide.
Figure~\ref{Fig:1}(c) shows the microscope image of the fabricated hybrid Si$_3$N$_4$ microresonator and AlN actuator device. 

We first characterize the strain-tuning of the microresonator resonances in the fundamental transverse-electric mode (TE$_{00}$). 
Figure~\ref{Fig:1}(d) shows the low-speed (DC) tuning curve, i.e. resonance frequency shift versus the applied voltage up to $\pm140$~V, of a TE$_{00}$ resonance at 1556~nm (see details in Supplementary Information). 
The average linear tuning coefficient at DC is $\delta\nu/\delta V=15.7$~MHz/V.
Note that, this resonance tuning is \emph{bi-directional}, in contrast to the uni-direction thermal tuning using heaters. 
A weak hysteresis is observed when cycling the voltage between $\pm140$~V.
While crystalline AlN is non-ferroelectric, such hysteresis is commonly caused by the trapped charges accumulated at AlN interfaces or combined with bulk defects\cite{Aardahl:99}.
Since the AlN actuator is capacitive, the operation current at high voltage is measured to be less than 2~nA, corresponding to less than 300~nW power consumption at 150~V voltage.
The calculated capacity based on the device geometry is 3.3~pF, corresponding to stored energy of 37~nJ at 150~V voltage.
Therefore, the main power consumption comes from the leakage current, likely caused by the defects and free charge carriers in polycrystalline AlN.

Figure~\ref{Fig:1}(e) shows the frequency tuning for all TE$_{00}$ resonances in the wavelength range from 1500~nm to 1630~nm,  calibrated using frequency-comb-assisted diode laser spectroscopy\cite{DelHaye:09, Liu:16}. 
The wavelength-dependent tuning rate results from the change of microresonator FSR ($\sim5$~MHz of 191~GHz FSR). 
The measured intrinsic quality factor, $Q_0>15\times10^6$,  with integrated AlN actuators is identical to bare microresonators without AlN\cite{Liu:18a} (see details in Supplementary Information), demonstrating that the monolithically integrated AlN actuators are compatible with ultralow-loss Si$_3$N$_4$ waveguide platform (linear optical loss of $\sim1$ dB/m).

\textbf{Voltage-controlled soliton microcombs}: 
We next demonstrate piezoelectric-assisted soliton initiation, switching and bandwidth tuning.  
A CW laser is coupled into the microresonator, and the soliton is initiated by strain-tuning the resonance to the laser\cite{Joshi:16} via varying the voltage, using the setup shown in Fig.~\ref{Fig:2}(a).
The laser is initially set blue-detuned by 1~GHz from a microresonator resonance, and launches 15~mW of power into the waveguide (60$\%$ coupling efficiency per chip facet).
A typical soliton step of millisecond duration is shown in Fig.~\ref{Fig:2}(b).
Though not required for soliton initiation, we monitor the resonance-laser detuning using an EOM and a vector network analyser (VNA)\cite{Guo:16}. 
As shown in Fig.~\ref{Fig:2}(c), the resonance is initially tuned to the laser (0 V$\rightarrow$81 V), and subsequently generates modulation instability (MI, 81 V) and a multi-soliton state (MS, 85 V).
Next, the AlN voltage is reduced such that the backward tuning enables switching\cite{Guo:16} to the single soliton state (SS, 79 V). 
The voltage is increased again (90 V) to increase the soliton bandwidth. 
Figure~\ref{Fig:2}(d) shows different soliton states with different applied voltages.
The voltage-controlled AlN actuator can be used to implement feedback and to eliminate detuning fluctuations for long-term soliton stabilization (see details in Supplementary Information). 

\begin{figure*}[t!]
\centering
\includegraphics{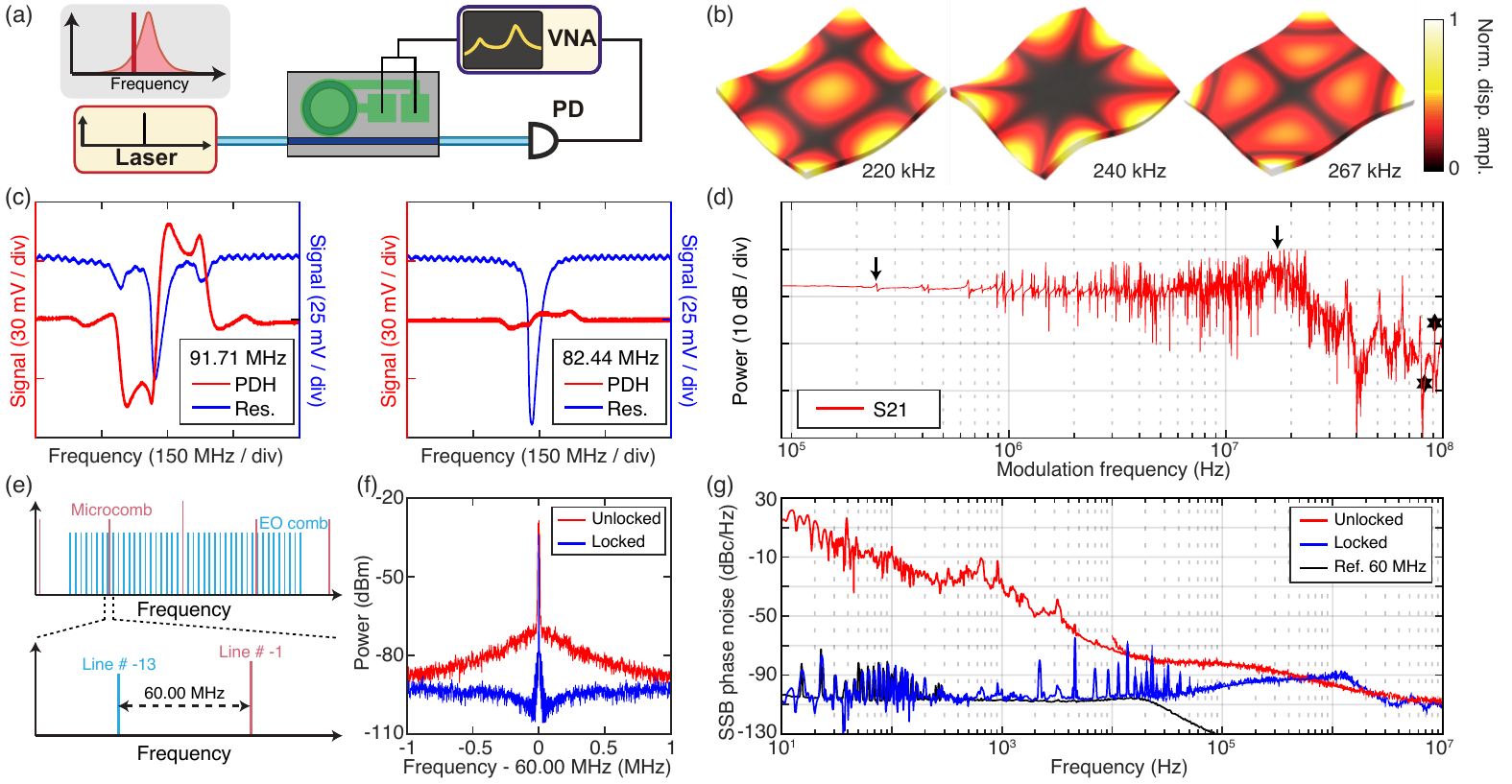}
\caption{
\textbf{High-speed piezoelectric actuation for on-chip PDH error signal generation and soliton repetition rate stabilization}. 
(a)~Experimental setup to measure $S_{21}(\omega)$. The laser frequency is set on the resonance slope.
(b)~Simulated contour modes of the photonic chip starting from 220 kHz. The color represents the displacement amplitude. 
(c)~Generated PDH error signals when modulating the AlN at 91.71 and 82.44~MHz, marked with stars in (d).
(d)~Electrical to optical signal transduction $S_{21}(\omega)$ of the AlN actuator. 
Arrows mark the contour mode at 246~kHz and the fundamental HBAR mode at $\sim17$~MHz. 
(e)~Schematic of soliton repetition rate stabilization using an EO comb with 14.6974~GHz line spacing. 
The $-1^{st}$ line of the microcomb and the $-13^{th}$ line of the EO comb are locked, referenced to a 60.0~MHz microwave signal. 
(f)~Measured beatnote signal of the $-1^{st}$ line of the microcomb and the $-13^{th}$ line of the EO comb, in the cases of locked and unlocked (free-running) states. Resolution bandwidth (RBW) is 1~kHz. 
(g)~Measured phase noise of the beat signal, in comparison to the 60.0~MHz microwave signal. 
The locking bandwidth of the AlN actuator is 0.6~MHz. 
}
\label{Fig:3}
\end{figure*}

\textbf{Fast soliton actuation and locking}:
Next, we show that the AlN actuator allows microresonator modulation which can be utilized to stabilize the soliton repetition rate.
Figure~\ref{Fig:3}(a) shows the experimental setup to characterize the frequency transduction $S_{21}(\omega)$ from the electrical to the optical domain. 
As shown in Fig.~\ref{Fig:3}(d), the measured $S_{21}(\omega)$ has multiple peaks extending from 200~kHz, which correspond to different mechanical modes of the photonic chip, excited by the AlN actuator.
The mechanical modes with frequency from 246~kHz to $\sim17$~MHz are contour modes\cite{Piazza:06} of the entire chip.
Figure~\ref{Fig:3}(b) shows the simulated contour modes around 246~kHz, using finite element modelling based on the actual chip size of $4.96\times4.96$~mm$^2$.
These contour modes only exist with free boundary conditions. 
The broad peak at $\sim17$~MHz is the fundamental HBAR mode\cite{Tian:19}, confined vertically over the chip thickness and material stack (8.4 $\mathrm{\mu}$m SiO$_2$ and 213 $\mathrm{\mu}$m Si).
The resonances at multiple of $\sim17$~MHz are higher-order HBAR modes. 
Though narrowband, the HBAR modes provide novel functionality, such as error signal generation for the Pound-Drever-Hall (PDH) technique. 
Figure~\ref{Fig:3}(c) compares the generated PDH error signals, when applying 82.44~MHz and 91.71~MHz modulation frequency directly on the AlN actuator (see details in Supplementary Information).
The error signal with 91.71~MHz frequency is $\times10$ stronger than that of 82.44~MHz frequency, since only the 91.71~MHz corresponds to an HBAR frequency, marked in Fig.~\ref{Fig:3}(d). 
More error signals at other frequencies are shown in Supplementary Information.

The relatively flat $S_{21}(\omega)$ response up to $\sim800$~kHz allows soliton repetition rate stabilization via AlN actuation.
As the soliton repetition rate, $\nu_\text{rep}=191$~GHz, is not directly measurable, we utilize an electro-optic frequency comb (EO comb) of 14.6974~GHz line spacing, and measure the beat signal between the $-1^{st}$ line of the microcomb and the $-13^{th}$ line of the EO comb, as illustrated in Fig.~\ref{Fig:3}(e).
Both microcomb and EO comb are generated with the same pump laser. 
The measured beat signal is further compared to a reference signal of 60.0~MHz, and the error signal is applied directly on the AlN actuator, such that the actuation on the microresonator stabilizes the soliton repetition rate to the EO comb line spacing (see details in Supplementary Information).
Figure~\ref{Fig:3}(f, g) compare the measured beat signal between the $-1^{st}$ line of the microcomb and the $-13^{th}$ line of the EO comb, and the phase noise of the beat signal in the cases of free-running and locked states, in comparison with the phase noise of the 60.0~MHz reference microwave.
The locking bandwidth, determined by the merging point of two phase noise curves, is 0.6~MHz.
Note that 0.6~MHz is a large bandwidth for piezoelectric optical frequency actuators, compared to conventional lasers where the piezo response is typically limited to a few kilohertz (similar to integrated heaters). 

\begin{figure*}[t!]
\centering
\includegraphics{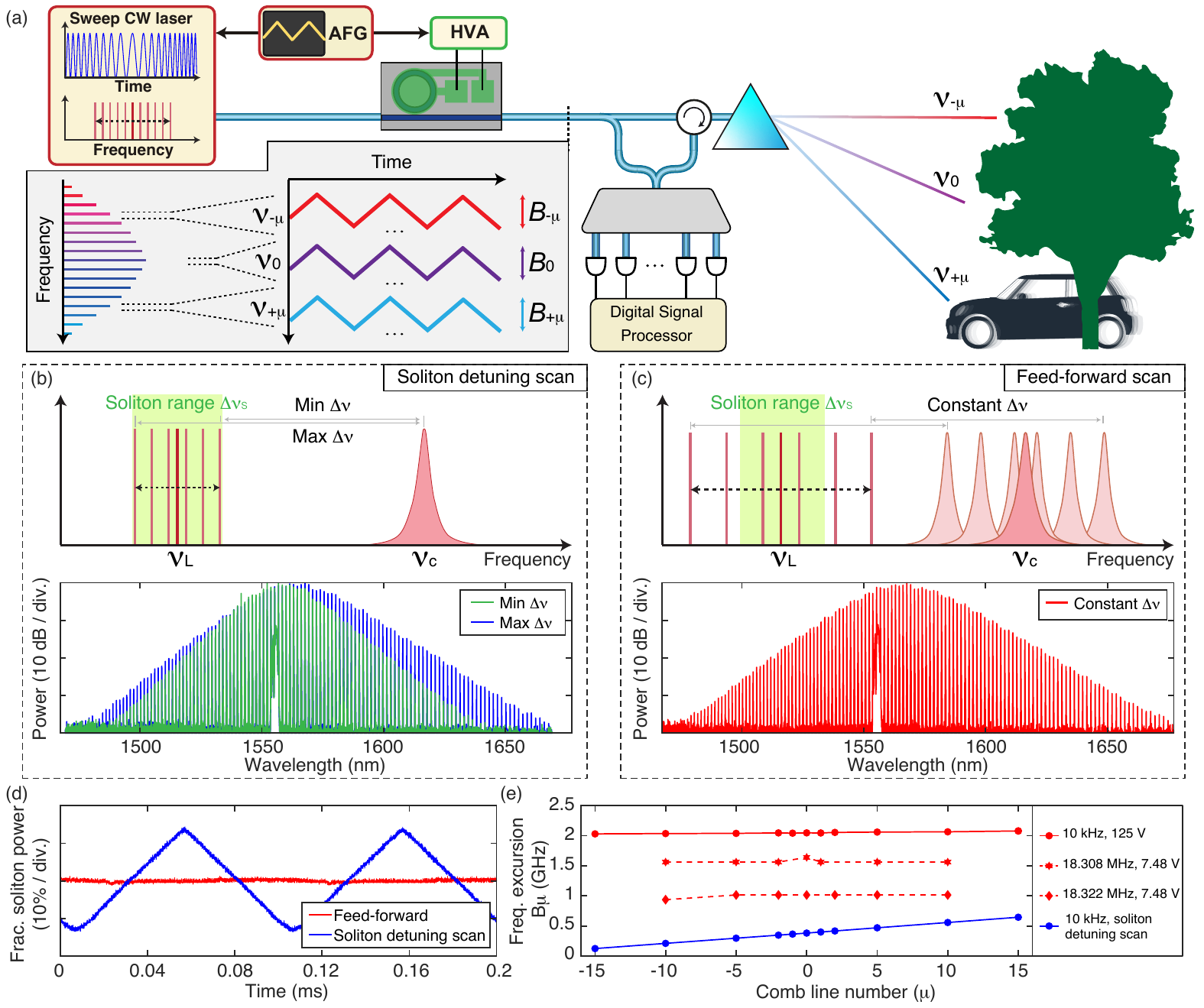}
\caption{
\textbf{Hybrid AlN-Si$_3$N$_4$ soliton LiDAR engine}.
(a)~Schematic of soliton-based parallel FMCW LiDAR. 
A chirped pump transduces synchronous modulation to other comb lines, with the same modulation rate and frequency excursion.
(b)~The scheme employing soliton detuning scan, with a fixed microresonator resonance $\nu_\text{c}$, and a pump centered at $\nu_\text{L}$ chirping within the soliton existence range (green shaded). 
By varying the soliton detuning $\Delta\nu$, the soliton spectrum varies correspondingly, as shown in the bottom. 
(c)~Feed-forward scheme employing synchronous modulation of the pump frequency $\nu_\text{L}$ and the resonance $\nu_\text{c}$, such that the soliton detuning $\Delta\nu$ is constant, also resulting in a constant soliton spectrum shown in the bottom. 
Note that in this scheme, the pump frequency excursion can be significantly larger than the soliton existence range (green shaded)
(d)~Comparison of both schemes for the integrated soliton pulse power. 
The soliton detuning scan shows a varying soliton pulse power, while the feed-forward scheme shows a constant power.
(e)~Comparison of the frequency excursions of different soliton comb lines, using different approaches.
}
\label{Fig:4}
\end{figure*}

\textbf{Soliton-based parallel LiDAR engine}:
Further, we show that the AlN actuator can be a key component for a soliton-based, parallel, frequency-modulated continuous-wave (FMCW) LiDAR\cite{Kuse:19, Riemensberger:19}. 
Previously demonstrated scheme\cite{Riemensberger:19} relies on scanning the pump laser's frequency $\nu_\text{L}$ over the soliton existence range $\Delta\nu_\text{s}=\nu_2-\nu_1$, i.e. $\nu_\text{c}-\nu_\text{L}\in[\nu_1, \nu_2]$, with $\nu_\text{c}$ being the resonance frequency, $\nu_1$ and $\nu_2$ being the boundary of soliton existence detuning range. 
This results in transferring the pump frequency chirp to all soliton comb teeth, as shown in Fig.~\ref{Fig:4}(a). 
Combined with diffractive optics that disperses multiple frequency lines, this approach to FMCW LiDAR allows high-speed parallel acquisition of both velocity and position in each pixel. 
However, the varying soliton detuning has several limitations \cite{Riemensberger:19}. 
First, the varying soliton detuning $\nu_\text{c}-\nu_\text{L}$ leads to variations in soliton spectrum bandwidth and power, which limits the number of usable optical channels.
Second, the Raman self-frequency shift\cite{Karpov:16, Yi:17} causes a varying soliton repetition rate (i.e. comb line spacing). 
Third, GHz frequency excursion of the pump laser ($\Delta\nu_\text{L}$) requires GHz-wide soliton existence range ($\Delta\nu_\text{s}$), necessitating an elevation in the pump power to several Watts. 

These limitations can be overcome using the AlN actuation on the microresonator, such that the microresonator resonance $\nu_\text{c}$ is \emph{modulated synchronously} to the pump frequency $\nu_\text{L}$, in order to maintain a constant soliton detuning $\nu_\text{c}-\nu_\text{L}$. 
We experimentally investigate both schemes, i.e. (a) scanning the pump laser within the soliton existence range, and (b) synchronously modulating the pump laser and the microresonator resonance (``feed-forward''), as shown in Fig.~\ref{Fig:4}(b, c).  
For the feed-forward scheme, a triangular signal of frequency $f_\text{mod}$ and peak-to-peak voltage $U_\text{pp}$ is applied on both the pump laser and the AlN actuator.
A phase shifter, an attenuator and a high-voltage amplifier (HVA) are used to modify the triangular signals in order to synchronize the resonance tuning to the laser tuning and to achieve the same frequency excursions of the laser and the resonance (see details in Supplementary Information). 
Figure \ref{Fig:4}(b, c, d) illustrate the differences in soliton spectrum and integrated soliton pulse power using both schemes, and highlight the advantages of the feed-forward scheme.
First, the pump frequency chirp $\Delta\nu_\text{L}$ can be significantly larger than the soliton existence range $\Delta\nu_\text{s}$, while simultaneously allowing soliton operation with tens of milliwatt of power, compatible with state-of-the-art integrated lasers\cite{HuangD:19, Boller:19}.
Second, the soliton spectrum and the repetition rate are nearly constant and are not affected by the pump frequency chirp. 

Figure~\ref{Fig:4}(e) compares the frequency excursions of different soliton comb lines.
The scheme where the soliton detuning is scanned (blue, solid line) at frequency $f_\text{mod}=10$ kHz, shows a frequency excursion dependence on the comb line number, resulting from the soliton repetition rate change $\Delta\nu_\text{rep}\approx$ 17.4~MHz per 0.383~GHz pump frequency excursion. 
Using the feed-forward scheme with $f_\text{mod}=10$ kHz and $U_\text{pp}=125$ V (red, solid line), a frequency excursion exceeding 2~GHz is achieved, which is more than $\times 5$  larger than the soliton existence range ($\Delta\nu_\text{s}\approx400$ MHz).
Moreover, the soliton repetition rate change is reduced to $\Delta f_\text{rep}\approx$ 0.066~MHz per 2.04~GHz pump frequency excursion.
More importantly, the HBAR and contour mechanical modes of our device enable modulation with higher frequency and efficiency (red, dashed lines).
By applying modulation with $f_\text{mod}=18.308$~MHz, coinciding with the fundamental HBAR mode and a contour mode, we impart 1.6 GHz pump frequency excursion with the voltage applied on the AlN of only $U_\text{pp}=7.48$ V, a CMOS-compatible voltage. 
This modulation corresponds to an equivalent tuning speed of 58.6 PHz/s. 
The corresponding resonance tuning coefficient is $\delta\nu/\delta V=219$~MHz/V, which is $\times 14$ larger than the DC tuning value of $\delta\nu/\delta V=15.7$~MHz/V.
Applying modulation with $f_\text{mod}=18.322$~MHz, while still coinciding with the fundamental HBAR mode but not a contour mode, the tuning coefficient is reduced to $\delta\nu/\delta V=133$~MHz/V.

In conclusion, we demonstrate integrated piezoelectric control of soliton microcombs, by monolithically integrating AlN actuators on ultralow-loss Si$_3$N$_4$ waveguides.
This novel capability not only benefits existing applications, but also allows synchronous scanning of the pump laser and photonic microresonator, as required for massively parallel FMCW LiDAR.
While polycrystalline AlN is used in our current work, the operation voltage can be reduced by more than $\times 2$ using scandium-doped AlN \cite{Umeda:13, Fichtner:19}. 
By future co-integration of CMOS electronic circuitry on a closeby die, compactly packaged soliton microcombs with rapid electronic actuation is attainable. 

\medskip
\begin{footnotesize}

\noindent \textbf{Funding Information}: This work was supported by Contract HR0011-15-C-055 (DODOS) from the Defense Advanced Research Projects Agency (DARPA), Microsystems Technology Office (MTO), by the Air Force Office of Scientific Research, Air Force Material Command, USAF under Award No. FA9550-15-1-0099, and by Swiss National Science Foundation under grant agreement No. 176563. (BRIDGE).

\noindent \textbf{Acknowledgments}: We thank Johann Riemensberger, Anton Lukashchuk and Maxim Karpov for the fruitful discussion.
E.L. acknowledge the support from the European Space Technology Centre with ESA Contract No. 4000116145/16/NL/MH/GM. The Si$_3$N$_4$ microresonator samples were fabricated in the EPFL center of MicroNanoTechnology (CMi), and Birck Nanotechnology Center at Purdue University.

\noindent \textbf{Author contributions}: J.L., H.T. and R.N.W. designed and fabricated the samples. J.L., H.T., J.H. and T.L. tested the samples and yield. E.L., A.S.R., J.L., G.L. and M.H.A. performed experiments. J.L., H.T., E.L., A.S.R. and G.L. analyzed the data. J.L., T.J.K. and S.A.B. wrote the manuscript, with the input from others. T.J.K. and S.A.B initiated and supervised the collaboration.

\noindent \textbf{Data Availability Statement}: The code and data used to produce the plots within this work will be released on the repository \texttt{Zenodo} upon publication of this preprint.
\end{footnotesize}
\bibliographystyle{apsrev4-1}
\bibliography{bibliography}
\end{document}


\title{Supplementary Information to:\\Monolithic piezoelectric control of soliton microcombs}

\author{Junqiu Liu}
\thanks{These authors contributed equally to this work.}
\affiliation{Institute of Physics, Swiss Federal Institute of Technology Lausanne (EPFL), CH-1015 Lausanne, Switzerland}

\author{Hao Tian}
\thanks{These authors contributed equally to this work.}
\affiliation{OxideMEMS Lab, Purdue University, 47907 West Lafayette, IN, USA}

\author{Erwan Lucas}
\thanks{These authors contributed equally to this work.}
\affiliation{Institute of Physics, Swiss Federal Institute of Technology Lausanne (EPFL), CH-1015 Lausanne, Switzerland}

\author{Arslan S. Raja}
\thanks{These authors contributed equally to this work.}
\affiliation{Institute of Physics, Swiss Federal Institute of Technology Lausanne (EPFL), CH-1015 Lausanne, Switzerland}

\author{Grigory Lihachev}
\affiliation{Institute of Physics, Swiss Federal Institute of Technology Lausanne (EPFL), CH-1015 Lausanne, Switzerland}

\author{Rui Ning Wang}
\affiliation{Institute of Physics, Swiss Federal Institute of Technology Lausanne (EPFL), CH-1015 Lausanne, Switzerland}

\author{Jijun He}
\affiliation{Institute of Physics, Swiss Federal Institute of Technology Lausanne (EPFL), CH-1015 Lausanne, Switzerland}

\author{Tianyi Liu}
\affiliation{Institute of Physics, Swiss Federal Institute of Technology Lausanne (EPFL), CH-1015 Lausanne, Switzerland}

\author{Miles H. Anderson}
\affiliation{Institute of Physics, Swiss Federal Institute of Technology Lausanne (EPFL), CH-1015 Lausanne, Switzerland}

\author{Wenle Weng}
\affiliation{Institute of Physics, Swiss Federal Institute of Technology Lausanne (EPFL), CH-1015 Lausanne, Switzerland}

\author{Sunil A. Bhave}
\email[]{bhave@purdue.edu}
\affiliation{OxideMEMS Lab, Purdue University, 47907 West Lafayette, IN, USA}

\author{Tobias J. Kippenberg}
\email[]{tobias.kippenberg@epfl.ch}
\affiliation{Institute of Physics, Swiss Federal Institute of Technology Lausanne (EPFL), CH-1015 Lausanne, Switzerland}

\maketitle

\renewcommand\thefigure{\thesection.\arabic{figure}}    
\setcounter{figure}{0} 

\section{Experimental setups}

\renewcommand\thefigure{\thesection.\arabic{figure}}    
\setcounter{figure}{0} 

\begin{figure*}[t!]
\centering
\includegraphics{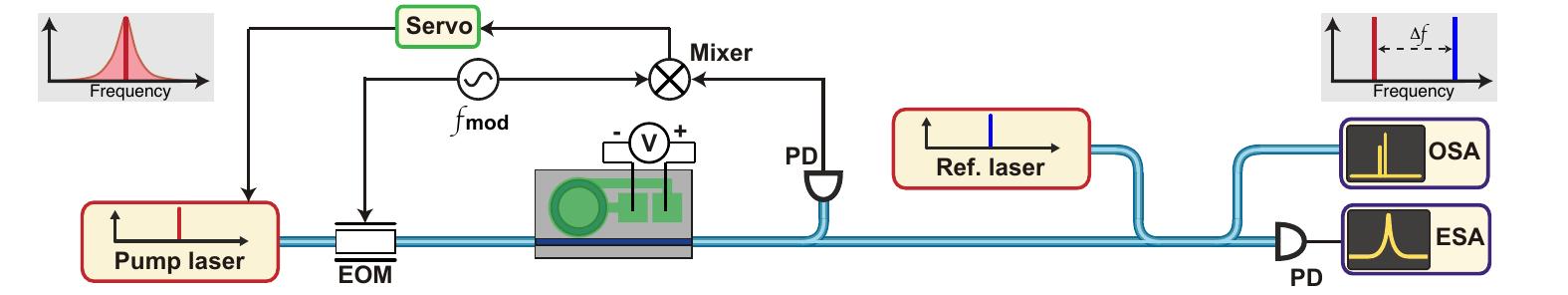}
\caption{
\footnotesize
Experimental setup to characterize the resonance tuning versus voltage applied. 
EOM: electro-optic modulator. 
OSA: optical spectrum analyzer. 
ESA: electrical spectral analyzer. 
PD: photodiode.
}
\label{Fig:SI1}
\end{figure*}

\noindent\textbf {Experimental setup to characterize resonance tuning}:
The experimental setup to characterize the resonance frequency tuning versus voltage applied on the AlN actuator is shown in Fig. \ref{Fig:SI1}. 
A tunable laser (Toptica CTL) is locked to a Si$_3$N$_4$ microresonator resonance, via a PDH lock loop using an EOM. 
When the resonance is tuned by varying the applied voltage, the laser frequency follows the resonance shift. 
The beat signal between the laser locked to the resonance and a reference laser (another Toptica CTL) is measured using a fast photodiode and an electrical spectrum analyzer (ESA). 
A programmable DC power supply (Keithley 2400) is used to apply the voltage on the AlN actuator. 
A ramp signal is applied on the power supply in order to output the voltage between $\pm140$ V with a voltage increment / decrement of 2.8 V. 
The interval time between two subsequent measurements is 200 ms. 
The change in the two lasers' beatnote signal recorded by the ESA corresponds to the resonance frequency shift, as one laser is locked to the resonance and the other is frequency-fixed. 
These measurements are repeated continuously for multiple (3 to 5) scans between $\pm140$ V in order to confirm the hysteresis.

\begin{figure*}[t!]
\centering
\includegraphics{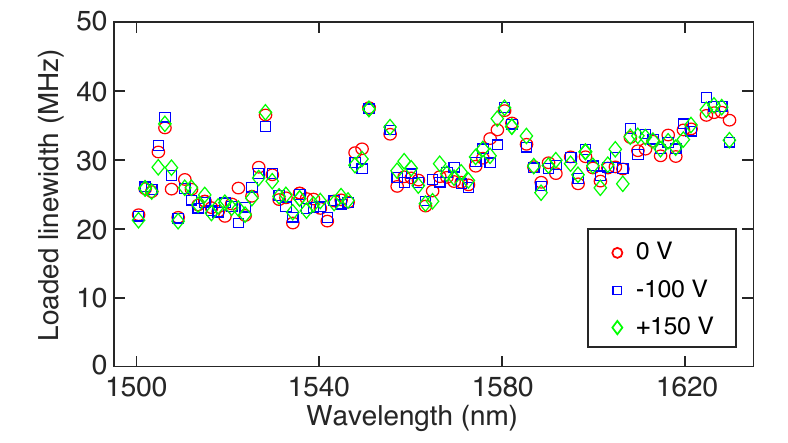}
\caption{
\footnotesize
Comparison of loaded linewidths with different applied voltages. 
No voltage-dependent linewidth change is observed.
}
\label{Fig:SI1b}
\end{figure*}

\noindent\textbf{Microresonator Q characterization results}:
Figure~\ref{Fig:SI1b} compares the measured loaded linewidths with different applied voltages.
The resonances remain critically coupled, and no linewidth change is observed.
The estimated intrinsic quality factor, $Q_0>15\times10^6$, with integrated AlN actuators is identical to bare microresonators without AlN\cite{Liu:18a}, demonstrating that the monolithically integrated AlN actuators are compatible with ultralow-loss Si$_3$N$_4$ waveguide platform (linear optical loss of $\sim1$ dB/m).

\begin{figure*}[t!]
\centering
\includegraphics{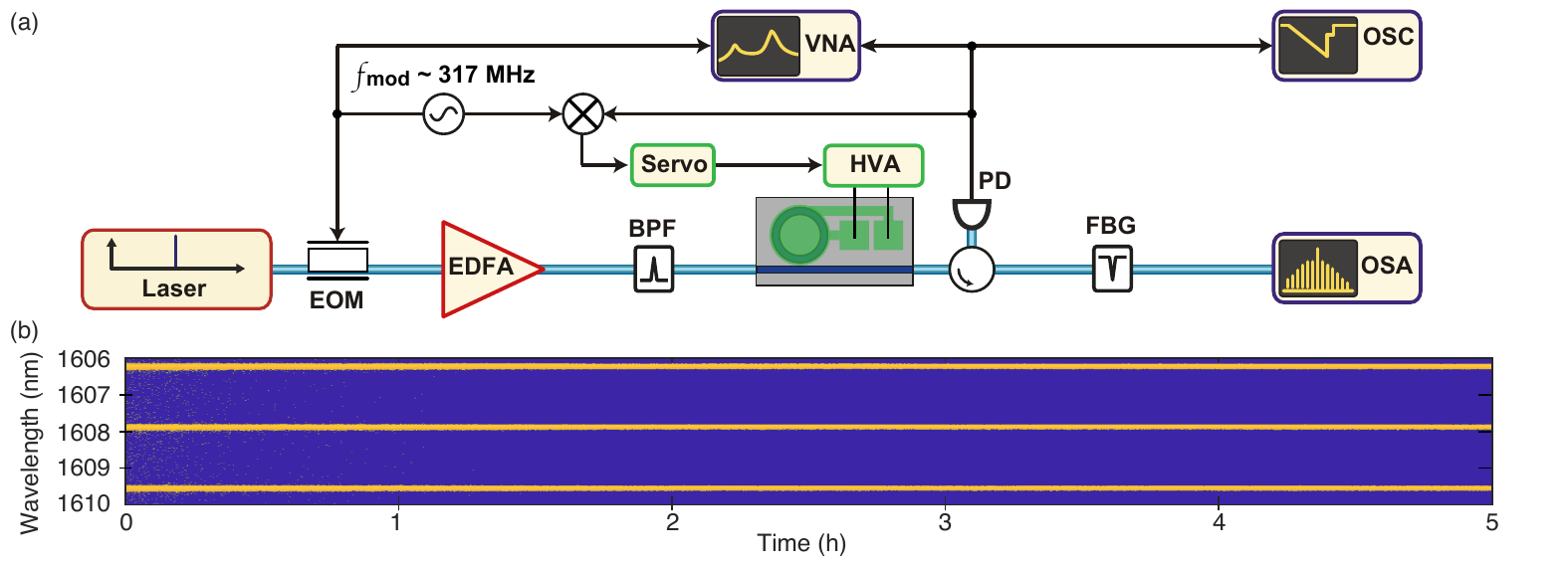}
\caption{
\footnotesize
(a)~Experimental setup for long-term stabilization of the soliton microcomb. 
OSC: oscilloscope. 
VNA: vector network analyzer. 
HVA: high voltage amplifier. 
BPF: bandpass filter. 
FBG: fiber Bragg grating. 
(b)~Soliton stabilization over 5 hour, realized by locking the resonance to the laser and maintaining the soliton detuning.
}
\label{Fig:SI1a}
\end{figure*}

\noindent\textbf{Experimental setup and result for Long-term stabilization of the soliton microcomb}: 
The experimental setup to stabilize the soliton microcomb over 5 hours is shown in Fig.~\ref{Fig:SI1a}(a). 
A feedback loop is applied in order to fix the soliton detuning at 317 MHz and eliminate the detuning fluctuation over a long term. 
The VNA is used only to monitor the soliton detuning over a long term.
Figure~\ref{Fig:SI1a}(b) shows the evolution of three soliton comb lines over 5 hours. 
The final soliton loss after 5 hours is caused by the drift of the fiber-chip coupling using suspended lensed fibers, and can be mitigated via gluing the fibers to the chip~\cite{Raja:20}. 

\begin{figure*}[t!]
\centering
\includegraphics{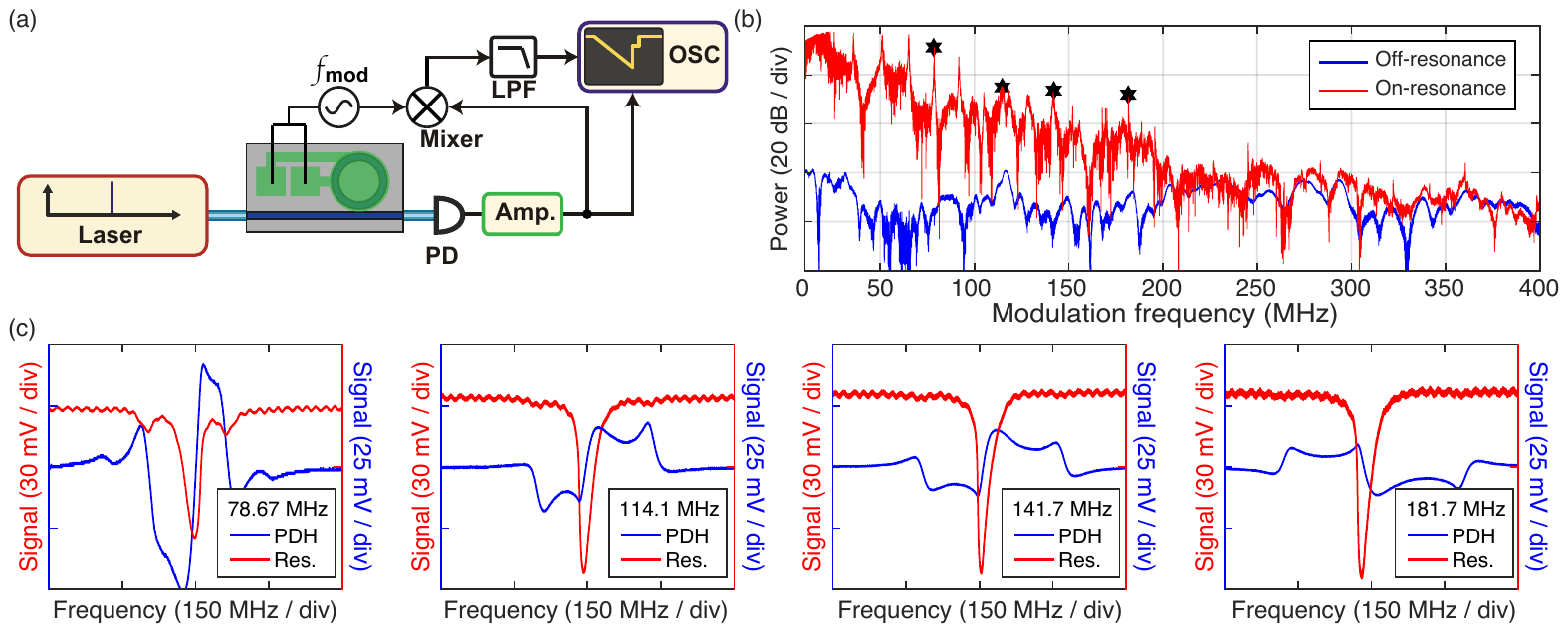}
\caption{
\footnotesize
On-chip generation of PDH error signals using the HBAR modes induced by the AlN actuation.
(a) Experimental setup.
LPF: low-pass filter. 
Amp.: RF power amplifier. 
(b) The measured $S_{21}(\omega)$ response of the AlN actuator in the linear frequency scale. 
Both cases, when the laser is on- and off-resonance, are measured.
(c) The PDH error signals modulated at the HBAR frequencies marked with stars in (b).
}
\label{Fig:SI2}
\end{figure*}

\noindent \textbf{Experimental setup to generate PDH error signals using HBAR modes}:
\indent Figure \ref{Fig:SI2}(a) shows the experimental setup to generate PDH error signals using HBAR modes induced by the AlN actuation. 
The measured $S_{21}(\omega)$ response of the AlN actuation, up to 400~MHz, is plotted in the linear frequency scale in Fig. \ref{Fig:SI2}(b), showing both cases when the laser is on- and off-resonance. 
Different modulation frequencies corresponding to different HBAR modes are investigated, which are marked with stars in Fig. \ref{Fig:SI2}(b). 
The PDH error signals modulated at these HBAR frequencies are shown in Fig. \ref{Fig:SI2}(c), as well as the studied microresonator resonance.
A microwave source providing $\sim8$~dBm RF power is used to modulate the Si$_3$N$_4$ microresonator via AlN actuation. 
The same RF power is used for modulation at all the HBAR frequencies.
The decrease in error signal contrast at higher HBAR frequency is likely caused by the lower acousto-optic transduction $S_{21}$.  

\section{Soliton microcomb source for parallel FMCW LiDAR}

\begin{figure*}[t!]
\centering
\includegraphics{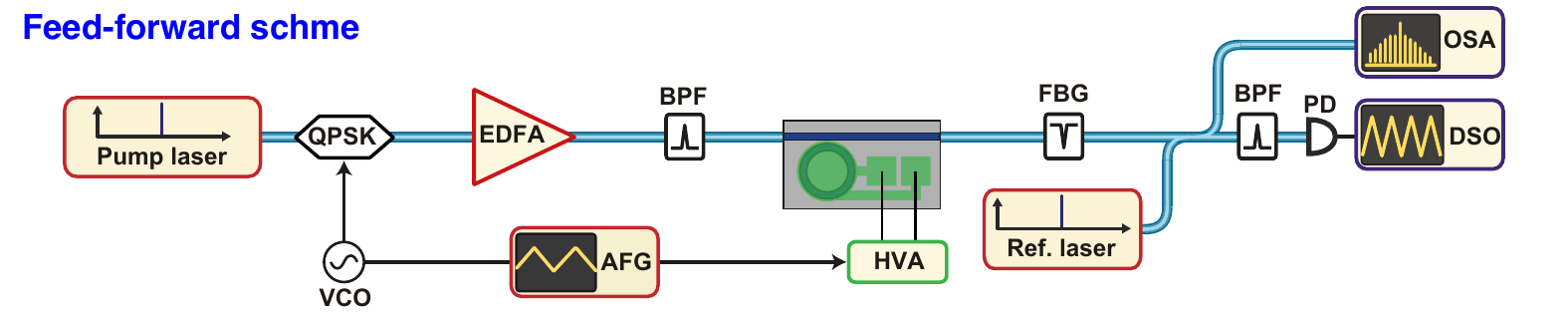}
\caption{
\footnotesize
Experimental setup for synchronous scan of the laser frequency and the microresonator resonance, using the feed-forward scheme. 
VCO: voltage-controlled oscillator. 
AFG: arbitrary waveform generator. 
QPSK: quadrature phase shift keying. 
DSO: digital storage oscilloscope.
}
\label{Fig:SI4}
\end{figure*}

Figure \ref{Fig:SI4} shows the experimental setup to synchronously scan the microresonator and the pump laser (i.e. the feed-forward scheme). 
A single-sideband modulator driven by a voltage-controlled oscillator (VCO) is used to fast scan the laser frequency, instead of directly scanning the laser piezo due to the limited piezo scan speed of our laser ($\sim200$ Hz). 
A voltage ramp signal from the same dual-channel arbitrary waveform generator (AWG) is applied on the VCO and on the AlN actuator. 
The ramp signal sent to the AlN actuator is further amplified by a high-voltage amplifier (HVA) with $\times50$ voltage amplification and 3-dB bandwidth of $\sim5$ MHz. 
The synchronous scan of the laser frequency and the microresonator resonance is performed by adjusting the amplitude and the phase of the ramp signal applied on the VCO.
A PDH lock can further improve the synchronization by locking the resonance to the laser with a constant frequency difference~\cite{Stone:18}.
Initially, a ramp signal from the AWG with a peak-to-peak voltage V$_\text{pp}$ of 3 V (HVA amplifies to 150 V) and 10 kHz scanning rate is applied on the AlN. 
The amplitude V$_\text{pp}$ and the phase of the ramp signal driving the VCO is adjusted until stable $\mathcal{C}.$-resonance is observed on VNA. 
The tuning into soliton states is realized either by changing the laser frequency via laser piezo tuning, or by turning on and off the VCO which allows fast tuning of the laser to the effectively red-detuned side of the resonance. 
A reference laser is used to probe the chirp of different comb lines (the pump line, $\pm 10^{th}$ comb lines etc). 
A fast oscilloscope of 2~GHz bandwidth and 5 GSamples/s is used to capture the heterodyne beatnote detected on the fast photodiode, for further off-line data processing such as fast Fourier transform and fitting triangular signal. 

\section{Soliton repetition rate stabilization using an EO comb}

\begin{figure*}[t!]
\centering
\includegraphics{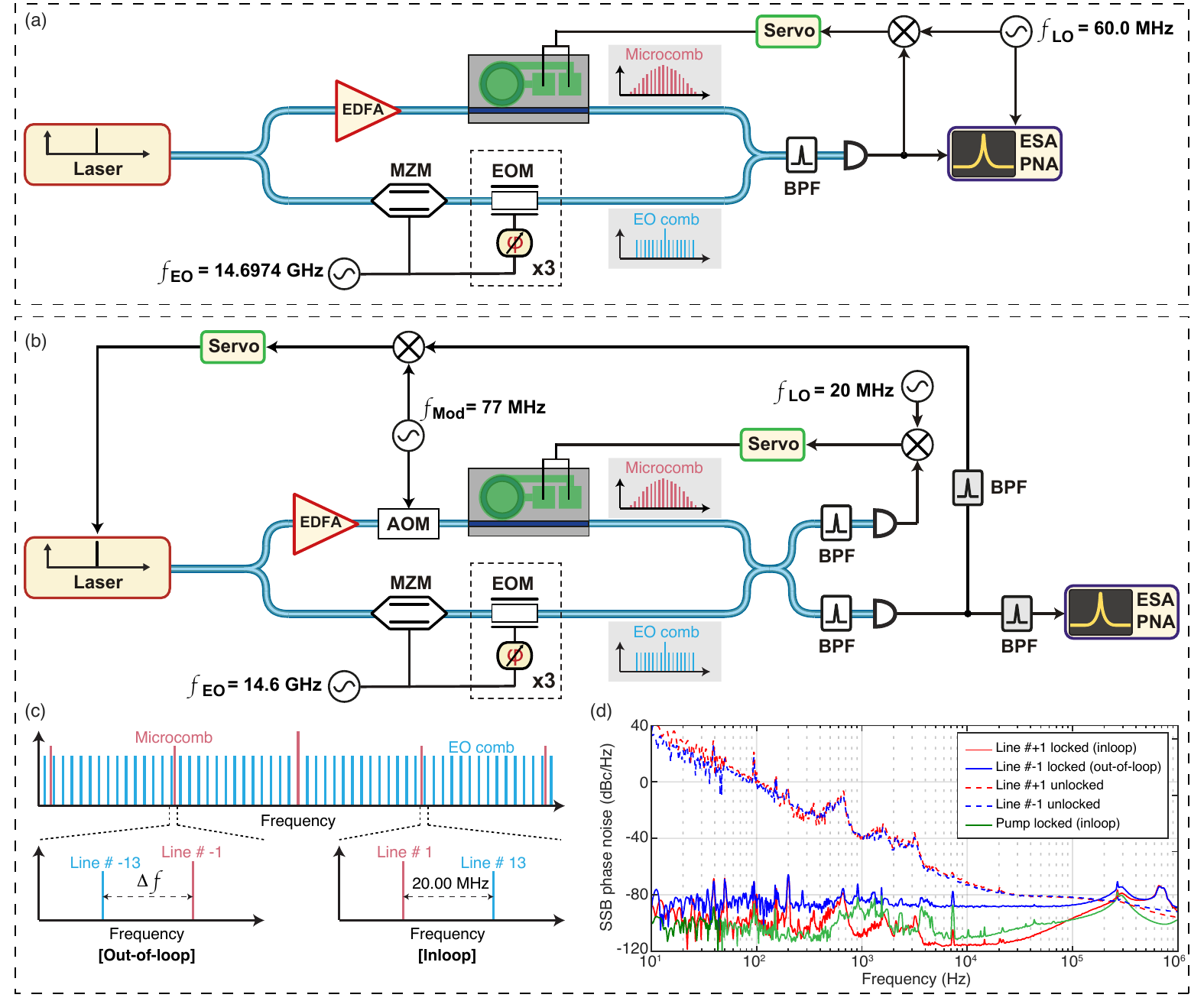}
\caption{
\footnotesize
Soliton repetition rate stabilization using an EO comb and AlN actuation. 
(a) Experimental setup to characterize the \emph{in-loop} phase noise of the beat signal between the $-1^{st}$ line of the microcomb and the $-13^{th}$ line of the EO comb. 
The beat signal and the phase noise of the beat signal are shown in Fig.~3(e, f) in the main manuscript.
(b) Modified experimental setup to characterize the \emph{out-of-loop} phase noise of the beat signal between the $-1^{st}$ line of the microcomb and the $+13^{th}$ line of the EO comb. 
(c) Schematic of referencing the microcomb to the EO comb. 
The beatnote (in-loop) between the $+1^{st}$ line of the microcomb and $+13^{th}$ line of the EO comb is detected and referenced to a 20.0~MHz microwave signal.
The beatnote (out-of-loop) between the $-1^{st}$ line of the microcomb and $-13^{th}$ line of the EO comb is characterized.
(d) Comparison of SSB phase noises measured in different cases.
MZM: Mach-Zehnder modulator. 
BPF: bandpass filter. 
PNA: phase noise analyzer.
}
\label{Fig:SI3a}
\end{figure*}

Figure \ref{Fig:SI3a}(a) shows the experimental setup to stabilize the soliton repetition rate referenced to an electro-optic frequency comb (``EO comb''). 
The EO comb is generated using a scheme described in Ref. \cite{Obrzud:17, Anderson:19}, and has a comb line spacing of 14.6974 GHz. 
The EO comb and soliton microcomb are pumped by the same laser (Toptica CTL). 
The measure the beat signal between the $-1^{st}$ line of the microcomb and the $-13^{th}$ line of the EO comb, is further compared to a reference signal of 60.0~MHz. 
The error signal is applied directly on the AlN actuator, such that the actuation on the microresonator stabilizes the soliton repetition rate to the EO comb line spacing.
The measured \emph{in-loop} phase noise of the beat signal between the $-1^{st}$ line of the microcomb and the $-13^{th}$ line of the EO comb, is shown in Fig.~3(g) in the main manuscript. 

To measure the \emph{out-of-loop} beat signal and its phase noise, we used a modified setup as shown in Fig.~\ref{Fig:SI3a}(b). 
The pump laser's frequency to generate the soliton microcomb is shifted by 77.0~MHz via a fiber-coupled acousto-optic modulator (AOM). 
The reason to shift the microcomb's pump frequency is to cancel out the drift and noise caused by the imbalanced paths in delayed self-homodyne measurement, by detecting the beatnote (77.0~MHz shift) between the pump lines of the microcomb and the EO comb. 
By down-mixing the 77.0~MHz heterodyne beatnote signal using the same microwave source that drives the AOM, the feedback signal is applied to the laser current such that the pump laser's frequency is stabilized and the noise in the delayed self-homodyne measurement is removed.

Then, the beatnote between the $+1^{st}$ line of the microcomb and $+13^{th}$ line of the EO comb is detected and referenced to a 20.0~MHz microwave signal, in order to stabilize the microcomb repetition rate. 
The entire schematic of referencing the microcomb to the EO comb is shown in Fig. \ref{Fig:SI3a}(c).
Figure \ref{Fig:SI3a}(d) compares the single-sideband (SSB) phase noise of the beat signals, for:
\begin{itemize}
  \item Dashed red: The free-running phase noise of the beat signal between the $+1^{st}$ line of the microcomb and the $+13^{th}$ line of the EO comb while the pump laser’s frequency is locked.
  \item Dashed blue: The free-running phase noise of the beat signal between the $-1^{st}$ line of the microcomb and the $-13^{th}$ line of the EO comb while the pump laser’s frequency is locked. 
  \item Solid green: When the pump laser's frequency is locked, the phase noise between the EO comb's pump and the microcomb's pump (shifted by 77~MHz). 
  \item Solid red: The \emph{in-loop}, locked phase noise of the beat signal between the $+1^{st}$ line of the microcomb and the $+13^{th}$ line of the EO comb.
  \item Solid blue: The \emph{out-of-loop} phase noise of the beat signal between the $-1^{st}$ line of the microcomb and the $-13^{th}$ line of the EO comb, when the $+1^{st}$ line of the microcomb and the $+13^{th}$ line of the EO comb are locked.
\end{itemize}
In the out-of-loop phase noise of the beat signal between the $-1^{st}$ line of the microcomb and the $-13^{th}$ line of the EO comb, a reduction in phase noise is observed with the AlN actuation. 
The locking bandwidth in this case is $>300$ kHz.

\begin{figure*}[t!]
\centering
\includegraphics{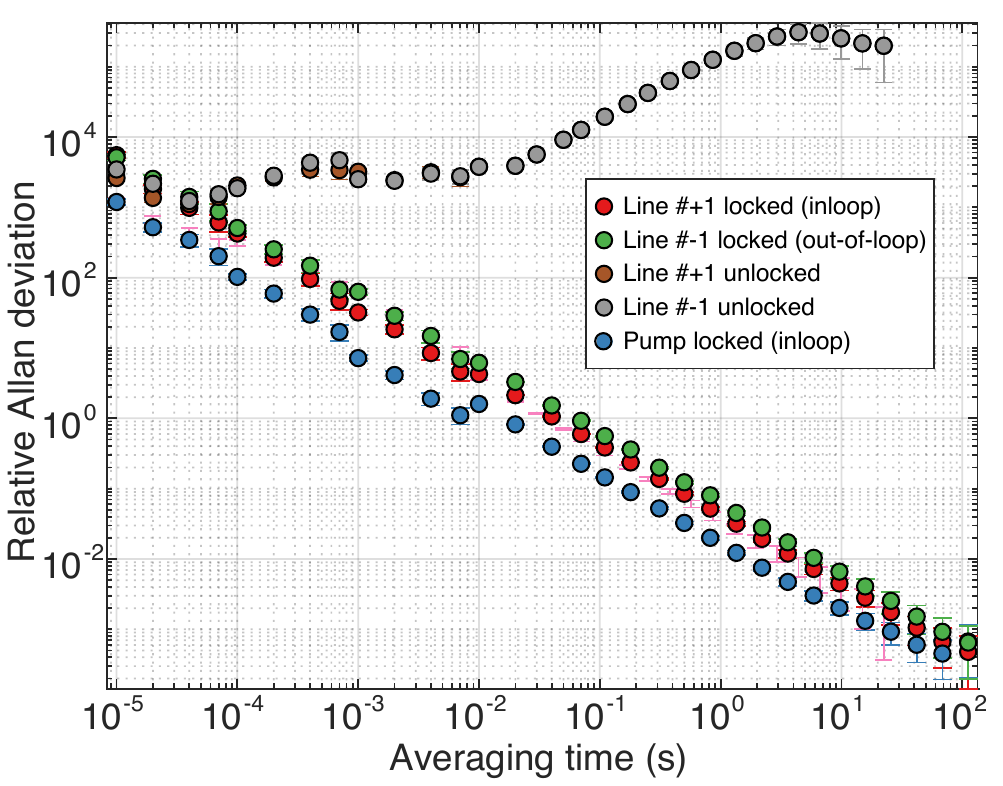}
\caption{
\footnotesize
Comparison of measured relative Allan deviation for the microcomb and EO comb beat signals in different cases, to evaluate the long-term stability of the locked system.
}
\label{Fig:SI3b}
\end{figure*}

To further evaluate the long-term stability of the locked system, frequency counting measurements of the relative Allan deviations are performed, as shown in Fig.~\ref{Fig:SI3b}.
The 77~MHz microwave source (used to lock the pump laser's frequency) is referenced to the 20 MHz microwave oscillator (used to down-mix the in-loop beat signal to derive the error signal). 
Similarly, the microwave source driving the EOMs (14.6974 GHz) for EO comb generation is also referenced to the same 20~MHz microwave oscillator. 
The relative Allan deviation of the beat signals between the free-running microcomb and EO comb are not converging, while the beat signal between the locked pump lines of the microcomb and EO comb show 10$^{-2}$ at 1 s averaging time.
After locking the soliton repetition rate to the EO comb by actuating on AlN, the in-loop and the out-of-loop beat signals show similar frequency stability. 

\bibliographystyle{apsrev4-1}
\bibliography{bibliography}